\documentclass[aps,prb,twocolumn,preprintnumbers,amsmath,amssymb,showpacs,superscriptaddress,10pt]{revtex4-1}
\usepackage{color}
\usepackage{graphics}
\usepackage{subfigure}
\usepackage{graphicx}
\usepackage{epsfig}
\usepackage{epstopdf}

\begin{document}
\allowdisplaybreaks
\setlength{\voffset}{1.0cm}
\title{Short-distance properties of Coulomb systems}
\author{Johannes Hofmann}
\email{j.b.hofmann@damtp.cam.ac.uk}
\affiliation{Department of Applied Mathematics and Theoretical Physics, University of Cambridge, Centre for Mathematical Sciences, Cambridge CB3 0WA, United Kingdom}
\author{Marcus Barth}
\email{marcus.barth@ph.tum.de}
\author{Wilhelm Zwerger}
\affiliation{Technische\,Universit{\"a}t\,M{\"u}nchen,\,Physik\,Department,\,
James-Franck-Strasse,\,85748 Garching,\,Germany}
\date{\today}
\begin{abstract}
We use the operator product expansion to derive exact results for the momentum distribution and the static structure factor at high momentum for a jellium model of electrons in both two and three dimensions. It is shown that independent of the precise state of the Coulomb system
and for arbitrary temperatures,  the asymptotic behavior is a power law in the momentum, whose strength is determined by the contact value of the pair distribution function $g(0)$. The power-law tails are quantum effects which vanish in the classical limit $\hbar \to 0$. A leading order virial expansion shows that the classical and the high-temperature limit do not agree.
\end{abstract}
\maketitle

\section{INTRODUCTION}

The basic constituents of ordinary matter are electrons and nuclei combined in such a way that there is no net overall charge. Within a non-relativistic approximation and treating the nuclei as
point particles, the interaction is fully described by an instantaneous Coulomb potential $\sim e^2/r$ at arbitrary distances. In spite of the long-range nature of this interaction and the divergent attractive force between electrons and nuclei at short distances, one expects an overall neutral Coulomb system to be stable in the sense that the ground state energy (or free energy at finite temperature) is finite and scales {\it linearly} with the total number $N$ of particles. It is one of the major accomplishments of theoretical physics to show that - beyond the exactly solvable case of the hydrogen atom - these expectations can indeed be proven rigorously. The proof crucially relies on the fact that electrons are fermions and are thus constrained by the Pauli principle~\cite{lieb10}. Since neither the size and mass nor the statistics of the nuclei play a role in this context, a simple approximation which captures much of the basic physics of Coulomb systems is the well-known jellium model, where the nuclei are treated as a homogeneous background that precisely cancel the negative charge of the Coulomb gas of electrons~\cite{pines66}. At zero temperature, this model is fully specified by the standard dimensionless interaction strength $r_s = r_0/a_0$. Here $a_0=\hbar^2/me^2$ is the Bohr radius while $r_0$ is the average spacing between electrons, connected with the electron density $n$ via $r_0 = (3/4\pi n)^{1/3}$ in 3D and $r_0 = (1/\pi n)^{1/2}$ in 2D, respectively.
Despite the fact that the periodic arrangement of the nuclei is ignored, the jellium model provides a reasonable starting point to describe elementary properties of metals like their cohesion energy or compressibility~\cite{pines66, fetter71, giuliani05}. Unfortunately, however, beyond the fundamental issue of stability and extensivity, there are hardly any exact results even for this highly simplified model. It is only in the high-density limit $r_s\ll 1$ where a perturbative expansion around the non-interacting Fermi gas is possible. A simple argument for this is provided by writing the jellium Hamiltonian  
\begin{align}
H &= - \frac{1}{r_s^2} \sum_i \nabla_i^2 + \frac{1}{r_s} \sum_{i < j} \frac{2}{|\mathbf{r}_i - \mathbf{r}_j|} + H_\mathrm{b}
\label{eq:1stquHamiltonian}
\end{align}
in dimensionless form, with Ry$=e^2/2a_0$ as the unit of energy and particle coordinates ${\bf r}_i$ measured in units of $r_0$ (both the energy $H_\mathrm{b}$ of the background as well as the interaction energy between electrons and the background are constants and thus need not be written explicitly). Clearly, as $r_s\to 0$, the kinetic energy dominates and the Coulomb interaction can be treated within perturbation theory. The expected ground state is a Fermi liquid, with a finite jump $0<Z<1$ of the momentum distribution at a spherical Fermi surface $|\mathbf{k}|=k_F$~\cite{footnote1}. While typical values of $r_s \approx 1-5$ in metals~\cite{ashcroft76} are outside the range of perturbation theory, at least the qualitative features of electrons in metals are captured correctly in this picture. For very large values of $r_s$, the uniform electron liquid is expected to eventually form a Wigner crystal, which minimizes the interaction energy in~\eqref{eq:1stquHamiltonian}. In addition, non-trivial phases such as anisotropic quantum liquid crystals are likely to appear at intermediate values of $r_s$. Indeed, in two dimensions a direct transition from a uniform electron liquid to a Wigner crystal as a function of $r_s$ can be ruled out by a quite general thermodynamic argument~\cite{jamei05}.  Moreover, even in the Wigner crystal, the electron spin gives rise to strong quantum fluctuations due to ring exchange processes, leading to a complex magnetic structure, see e.g. Refs.~\cite{chakravarty99, bernu01,katano02}.

In view of the still poorly understood phase diagram of even the simplest model for a many-body system with Coulomb interactions, it is of considerable interest to derive exact relations that hold independent of the interaction strength and   the particular state in question.  Our aim in the present work is to show that such relations follow directly from the operator product expansion (OPE) of quantum field theory. They constrain the short-distance physics of Coulomb systems in a manner which is analogous to the so-called Tan relations~\cite{tan05, braaten12, castin12} for fermions with interactions that have effectively zero range. Physically, these relations rely on the fact that at sufficiently short distances only the two-body interaction is relevant. At the level of the many-body ground state wave function for Coulomb systems, this observation goes back to Kimball~\cite{kimball73,kimball75,rajagopal77}.  Beyond providing a concise derivation of these relations, a novel and quite non-trivial aspect of our present derivation via the OPE is the fact that the relations will be shown to be valid at the level of {\it operators}. As a result, they apply to {\it any} state of the system, e.g., to a Fermi liquid or a Wigner crystal,  at zero or at finite temperature and also in a few-body situation. The only change is the value of the 'contact' $g(0)$, a dimensionless measure of the probability that two electrons with
opposite spin are found at a coincident point in space.

In order to illustrate the universal features in the short-distance behavior of Coulomb systems, it is instructive to consider the hydrogen atom as a simple and exactly solvable system.
In fact, the basic relations that will subsequently be proven for the many-body case
show up already in this elementary textbook problem~\cite{footnote2}.
The wave function for relative motion in the hydrogen atom has the well-known form~\cite{gottfried66}
\begin{align}
\nonumber \psi_{nlm} (\mathbf{r}) = & \left(\frac{2}{a_0 n}\right)^{3/2} \sqrt{\frac{(n-l-1)!}{2 n (n+l)!}} Y_{lm} (\theta ,\varphi ) \\ & \times \left(\frac{2 r}{a_0 n}\right)^l  e^{-\frac{r}{a_0 n}} L_{n-l-1}^{2 l+1}\left(\frac{2 r}{a_0 n}\right), \label{eq:hydrogen-wavefunction}
\end{align}
where $Y_{lm} (\theta ,\varphi )$ and $L_{n-l-1}^{2 l+1} (x)$ are spherical harmonics and generalized Laguerre polynomials, respectively (we assume the proton to be infinitely heavy, so the reduced mass is equal to the electron mass $m$). Its Fourier transform $\Upsilon_{nlm} (\mathbf{q})$ has been calculated by Podolsky and Pauling~\cite{podolsky29}. It is given by
\begin{align}
\nonumber \Upsilon_{nlm} (\mathbf{q}) = & 2^{2 l+4} \pi (a_0 n)^{3/2} (-i)^l l!  \sqrt{\frac{n (n-l-1)!}{(n+l)!}} Y_{lm}(\vartheta ,\phi ) \\ & \times \frac{\zeta ^l}{\left(\zeta ^2+1\right)^{l+2}} \, C_{n-l-1}^{l+1}\left(\frac{\zeta ^2-1}{\zeta ^2+1}\right),
\label{eq:hydrogen-momentumwf}
\end{align}
where $\zeta = q \, a_0 n$, and $ C_{n-l-1}^{l+1}(x)$ denotes a Gegenbauer polynomial. The momentum distribution of the electron is the absolute square of the momentum space wave function. Using Eq.~\eqref{eq:hydrogen-momentumwf}, $|\Upsilon_{nlm} (\mathbf{q})|^2$ turns out to decrease asymptotically as $1/q^{8+2l}$ for large momentum $\zeta = q \, a_0 n\gg 1$. The leading order term
\begin{align}
|\Upsilon_{n00}(\mathbf{q})|^2 &= \left( \frac{8\pi}{a_0} \right)^2 \frac{|\psi_{n00}(0)|^2 }{q^8} + \mathcal{O}(1/q^{10})
\label{eq:hydrogen-md-tail}
\end{align}
in the momentum distribution therefore only involves the contribution from $s$-states.  They are the only ones with a finite probability density $|\psi_{n00}(0)|^2 = 1/ \pi (a_0 n)^3$ for the electron and proton to be found at a coincident point in space.

Remarkably, the same 'contact' density also appears in the high-momentum tail of the atomic form factor $\rho_{nlm}(\mathbf{q})$, which is the Fourier transform of the electronic density distribution.  Its leading contribution at large $q$
\begin{align}
\label{eq:form-factor}
\rho_{n00}(\mathbf{q}) &= \int d^3 r\, e^{-i \mathbf{q} \cdot \mathbf{r}} |\psi_{n00} (\mathbf{r})|^2 \nonumber \\
&= \frac{16 \pi}{a_0} \frac{|\psi_{n00}(0)|^2}{q^4} + \mathcal{O}(1/q^5)
\end{align}
comes again from $s$-states, while higher angular momenta are associated with faster decaying power laws. As for the momentum distribution, the coefficient of the high-momentum tail contains the contact density $|\psi_{n00}(0)|^2$. Moreover, both the momentum distribution and the form factor depend only on the magnitude $q=|\mathbf{q}|$ of the wave vector, i.e., they have spherically symmetric tails since only $s$-states contribute. As will be shown in the following, the power laws found in the hydrogen atom and the fact that the physics at short distances is rotation invariant also show up in the many-body context, even for inhomogeneous or anisotropic phases.  More precisely, the momentum distribution is replaced by the Fourier transform of the one-particle density matrix, while the atomic form factor becomes the static structure factor $S(\mathbf{q})$ of the many-body system.

The article is structured as follows: in Sec.~\ref{sec:methods}, we introduce the jellium model, the one- and two-particle density matrix as well as some basics of the operator product expansion. In Sec.~\ref{sec:wavefunction}, the OPE is used to derive the exact short-distance behavior of general many-body wave functions. Moreover, it is shown that this implies
power-law tails in both the momentum distribution and the static structure factor which depend on the particular state in question only through the value of the contact $g(0)$. A direct computation of the short-distance OPE of the density-density correlator and the one-particle
density matrix is presented in Sec.~\ref{sec:asymptotics}. Finally, an explicit calculation of the contact in both the classical and the high-temperature limit is given in Sec.~\ref{sec:hight}. It is shown that for a Coulomb system these limits give opposite results and thus are not equivalent. The article is concluded by a summary and outlook, Sec.~\ref{sec:summary}. The Feynman rules of the diagrammatic calculation and some details of the evaluation of some few-particle matrix elements are discussed in Apps.~\ref{sec:feyrules} and~\ref{sec:diagcalc}.

\section{JELLIUM AND OPE}\label{sec:methods}

In second-quantized form, the Hamiltonian of the jellium model is given by
\begin{align}
H &= H_\mathrm{b} + \int d^dx \, \psi_\sigma^\dagger \frac{- \hbar^2 \nabla^2}{2m} \psi_\sigma^{} (\mathbf{x}) \nonumber \\
&+ \frac{1}{2} \int d^dx \int d^dx' \, \psi_{\sigma}^\dagger(\mathbf{x}) \psi_{\sigma '}^\dagger(\mathbf{x}') \frac{e^2}{|\mathbf{x} - \mathbf{x}'|} \psi_{\sigma '}^{}(\mathbf{x}') \psi_{\sigma}^{}(\mathbf{x}) , \label{eq:Hamiltonian}
\end{align}
where a summation over spin indices $\sigma = \uparrow,\downarrow$ is implied. Since the Hamiltonian involves only one- and two-body interactions, the expectation value of the energy in a state described by an arbitrary $N$-body density matrix only involves the reduced one- and two-particle density matrices $\gamma^{(1)}$ and $\gamma^{(2)}$~\cite{lieb10}. In a spin-resolved form and in a position space representation, the former can be expressed as
\begin{align}
\gamma^{(1)}_\sigma({\bf x}, \mathbf{x}') &= \langle \psi_\sigma^\dagger({\bf x}) \psi_\sigma^{}(\mathbf{x}') \rangle\, .
\label{eq:onepdm-def}
\end{align}
Its Fourier transform with respect to $\mathbf{x}-\mathbf{x}'$ then gives the momentum distribution (see Eq.~\eqref{eq:momentumdist} below). Regarding the two-particle density matrix
$\gamma^{(2)}(1,1'; \, 2,2')$, one only needs the diagonal elements $1\!=\!1'$, $2\!=\!2'$, which define a dimensionless, spin-resolved pair distribution function
\begin{equation}
n_{\sigma}(\mathbf{x})\, n_{\sigma'}(\mathbf{x}')\, g_{\sigma,\sigma'} (\mathbf{x}, \mathbf{x}')
= \left< \psi_{\sigma}^{\dagger} (\mathbf{x}) \psi_{\sigma'}^{\dagger} (\mathbf{x}') \psi_{\sigma'}^{} (\mathbf{x}') \psi_{\sigma}^{} (\mathbf{x}) \right>.
\label{eq:PD-definition}
\end{equation}
The pair distribution function is a measure of the probability to find an electron with spin projection $\sigma'$ at position $\mathbf{x}'$ given an electron with spin projection $\sigma$ is at $\mathbf{x}$. For a completely uncorrelated system one has $g(\mathbf{x}, \mathbf{x}')\equiv 1$. Note that there is no assumption here about translation invariance, which is certainly broken in the Wigner crystal. The total pair distribution function
\begin{align}
g(\mathbf{x}, \mathbf{x}') &= \sum_{\sigma, \sigma'} \frac{n_\sigma({\bf x}) n_{\sigma'}({\bf x}') g_{\sigma \sigma'}({\bf x}, {\bf x}')}{n({\bf x}) n({\bf x}')}
\label{eq:g_2total}
\end{align}
of a spin one half Fermi gas is a weighted sum of contributions $g_{\sigma,\sigma'}$. They all approach unity as $\mathbf{x}-\mathbf{x}'\to\infty$ and so does $g(\mathbf{x}, \mathbf{x}')$.  For small separations $\mathbf{x}- \mathbf{x}' \to 0$, in turn, the pair distribution function for equal spins vanishes quadratically because of the Pauli principle. Taking into account the possibility of a non-vanishing spin polarization $\zeta=(n_{\uparrow}-n_{\downarrow})/n$, one finds
\begin{equation}
g(0)= \frac{1}{2}\left( 1-\zeta^2\right)\, g_{\uparrow\downarrow}(0)
\label{eq:g_2onsite}
\end{equation}
for the total pair distribution function at vanishing separation in the translationally invariant case. Note that in a situation where the electronic state is not translation invariant, the local value $g(0)=g(\mathbf{R}, 0)$ of the pair distribution function depends also on the 'center-of-mass' coordinate $\mathbf{R}=(\mathbf{x}+\mathbf{x}')/2$, a dependence which is suppressed in the following.   

Both the one-particle density matrix and the pair distribution function can be expressed as expectation values of operators at different points in space. The operator product expansion - specified here to the relevant case of equal times - provides an expansion of an operator product ${\cal O}_a\, {\cal O}_b$ at nearby points in space in terms of local operators:
\begin{equation}
{\cal O}_a(\mathbf{R} - \frac{\mathbf{r}}{2}) {\cal O}_b(\mathbf{R} + \frac{\mathbf{r}}{2}) = \sum_{n} W_n({\bf r}) {\cal O}_n({\bf R}) .
\label{eq:opedef}
\end{equation}
It is important to emphasize that Eq.~\eqref{eq:opedef} is an operator relation, i.e., it is valid for expectation values between {\it any} state. The state-independent coefficients $W_n({\bf r})$ are ordinary c-numbers and are called the Wilson coefficients. They depend both on $n$ and the specific operators ${\cal O}_a$ and ${\cal O}_b$ which appear on the left-hand side of Eq.~\eqref{eq:opedef}. The scaling dimension $\Delta_n$ of a local operator ${\cal O}_n$ that contains $N_{\cal O}$ fermion creation and annihilation operators is defined by the property that the correlation between ${\cal O}_n$ and its hermitian conjugate at points separated by a small distance $r$ and time $t$ asymptotically scales as $t^{-\Delta_n} \exp\left[-i N_{\cal O} m r^2/2t\right]$. For example, the operator $\psi_\sigma^\dagger $ has scaling dimension $\Delta=d/2$. The values of $\Delta_n$ determine the dependence of the Wilson coefficients at small separation ${\bf r}$ via
\begin{align}
W_n({\bf r}) &= r^{\Delta_n - \Delta_a - \Delta_b} \, f(r/a_0, \hat{{\bf e}}_{\bf r}) ,
\end{align}
where $f$ is a function of the dimensionless ratio $r/a_0$ and the unit vector $ \hat{{\bf e}}_{\bf r}$, which reflects a possible angular dependence. The operators ${\cal O}_n$ with the lowest scaling dimension therefore govern the behavior of an operator product at small separation. In particular, Wilson coefficients which are non-analytic in $r$ give rise to power law tails of the associated correlator ${\cal O}_a {\cal O}_b$ at large momentum.

Regarding the question whether the OPE \eqref{eq:opedef} is a convergent rather than an asymptotic expansion, precise statements have only been given in the context of relativistic~\cite{wilson72} and, in particular, conformal field theories. In the latter case the OPE can be shown to have infinite radius of convergence~\cite{mack77,pappadopulo12}. For non-relativistic quantum field theories, like in our present problem, mathematically precise results on the convergence of the OPE are unfortunately not available. The OPE for the specific case of Coulomb systems may however be justified a posteriori by the fact that our main results like the short distance behavior of the many body wave function \eqref{eq:shortmanyp} and the cusp condition \eqref{eq:OPE-dd-correlator} agree with results derived in a mathematically precise manner via different methods~\cite{hoffmann-ostenhof92}.

In practice, the Wilson coefficients may be determined by performing few-particle calculations. Indeed, since they are state-independent, it is sufficient to calculate the matrix element of Eq.~\eqref{eq:opedef} between simple (few-particle) states for which $\langle {\cal O}_n \rangle \neq 0$. The coefficients $W_n({\bf r})$ then follow by matching both sides of Eq.~\eqref{eq:opedef}. As will be shown below, an operator of particular interest in the present context is the two-particle operator
\begin{equation}
{\cal O}_c(\mathbf{R})=\psi_\uparrow^\dagger   \psi_\downarrow^\dagger \psi_\downarrow^{}\psi_\uparrow^{}(\mathbf{R})\, .
\label{eq:contact}
\end{equation}
In analogy to the notion used for fermions with short range interactions, we shall refer to this as the contact operator. It has a finite expectation value in the presence of Coulomb interactions and thus the scaling dimension $\Delta_c=2d$ of the contact operator is the one inferred from simple dimensional analysis. This is quite different from the case where the interactions have zero range and $\mathcal{O}_c$ acquires an anomalous dimension two~\cite{braaten08}. The contact is a central quantity which determines the leading short-distance singularities of Coulomb systems and, in particular, the magnitude of the high-momentum tails of both the momentum distribution and the structure factor. In a translation invariant situation, the contact is equal to $n^2$ times the local value of the pair distribution function $g(0)$. Before proceeding to derive these results from the OPE in explicit form, we note that our derivation remains unchanged if the sign of the interaction is reversed. All the results of this article can thus be extended to the case of an attractive Coulomb interaction by simply changing $e^2 \to - e^2$.

\section{OPE FOR THE MANY-BODY WAVEFUNCTION}\label{sec:wavefunction}

\begin{figure}[t!]
\begin{center}
\subfigure[]{\scalebox{0.7}{\epsfig{file=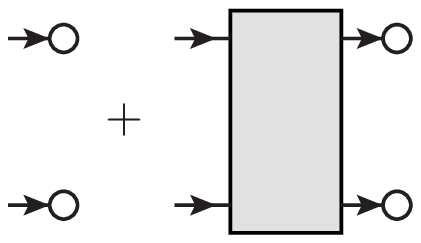,bb=231 646 364 721}}} \hspace*{0.5cm}
\subfigure[]{\scalebox{0.7}{\epsfig{file=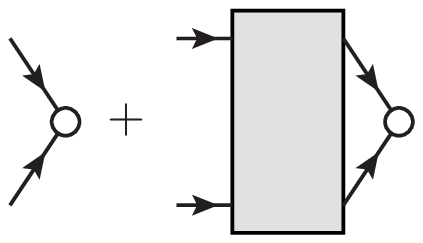,bb=231 646 364 721}}}
\caption{(a) Contribution to the two-particle matrix element of the operator $\psi(- \tfrac{\mathbf{r}}{2}) \psi(\tfrac{\mathbf{r}}{2})$, denoted by the white circles. (b) Same for the operator $\psi \psi(0)$. The T-matrix is denoted by a gray rectangle.}
\label{fig:1}
\end{center}
\end{figure}

The crucial physical insight, already implicit in the work of Kimball~\cite{kimball73,kimball75,rajagopal77}, relies on the intuition that the many-body wavefunction factorizes into a two-body contribution and a remainder whenever two particle coordinates come closer than the average interparticle distance. In this limit, the two particles only feel their mutual Coulomb repulsion at short distance, with negligible corrections from the medium. This type of argument has in fact been used by various authors~\cite{zhang09, werner12} in the derivation of the Tan relations for Fermi gases with short range interactions. In order to prove the validity of this intuitive picture, we use the operator product expansion for the special case of the operator
\begin{align}
{\cal O}_a({\bf x}) {\cal O}_b({\bf y}) = \psi_\uparrow({\bf x}) \psi_\downarrow({\bf y}) .
\end{align}
Inserting the general form~\eqref{eq:opedef} of the OPE,  the $N$-particle wavefunction $\Psi$ corresponding to an arbitrary $N$-particle state $|\Psi_N\rangle$ can be expanded as a formal power series
\begin{align}
&\Psi(- \frac{\mathbf{r}}{2}, \uparrow; \frac{\mathbf{r}}{2}, \downarrow; \mathbf{r}_3, \sigma_3; \ldots) \nonumber \\
&= \frac{1}{\sqrt{N_\uparrow! N_\downarrow!}} \, \langle 0 | \psi_\uparrow(- \frac{\mathbf{r}}{2}) \psi_\downarrow(\frac{\mathbf{r}}{2}) \prod_{l=3}^N \psi_{\sigma_l}({\bf r}_l) | \Psi_N \rangle \nonumber \\
&= \sum_n W_n({\bf r}) \, \frac{1}{\sqrt{N_\uparrow! N_\downarrow!}} \, \langle 0 | {\cal O}_n({\bf 0}) \prod_{l=3}^N \psi_{\sigma_l}({\bf r}_l) | \Psi_N \rangle ,
\label{eq:wavefunctionfac}
\end{align}
where $W_n({\bf r})$ are the Wilson coefficients in an OPE of the operator $\psi_\uparrow({\bf x}) \psi_\downarrow({\bf y})$, which can be written as the sum of a spin singlet operator
\begin{equation}
\psi({\bf x}) \psi({\bf y})=\frac{1}{2} (\psi_\uparrow({\bf x}) \psi_\downarrow({\bf y}) - \psi_\downarrow({\bf x}) \psi_\uparrow({\bf y}))
\label{eq:singlet}
\end{equation}
and a triplet operator which is symmetric in the spin indices. The leading order term in the OPE is associated with the operator $\psi \psi(0)$, whereas a similar contribution of the triplet operator vanishes since the fermion fields anticommute. Both singlet and triplet operators contribute in higher orders involving additional derivatives. The Wilson coefficient of the leading order can be obtained by taking the expectation value of Eq.~\eqref{eq:singlet} between the vacuum and a two-particle state with (on-shell) energy $p^2/m$. The corresponding diagrams are sketched in Fig.~\ref{fig:1}. (A brief summary of  the Feynman rules in momentum space for the jellium Hamiltonian~\eqref{eq:Hamiltonian} is given in App.~\ref{sec:feyrules}.)
We can express this matrix element as a one-body scattering wave function
\begin{align}
&\langle 0 | \psi(- \frac{\mathbf{r}}{2}) \psi(\frac{\mathbf{r}}{2}) | {\bf p}, \uparrow; - {\bf p}, \downarrow \rangle \nonumber \\
&= \langle {\bf r} | 1 + G_0 T | {\bf p}, \uparrow; - {\bf p}, \downarrow \rangle = \psi_p({\bf r}) ,
\end{align}
using the bare retarded two-particle propagator $G_0$ and the T-matrix $T$. In the second line, we have used the Lippmann-Schwinger equation for the scattering wavefunction $\psi_p({\bf r})$ with energy $p^2/m$. Similarly, the matrix element of the operator $\psi \psi(0)$ is $\psi_p(0)$~\cite{footnote3}. We determine its Wilson coefficient by matching this matrix element to leading order in the energy $p^2/m$ of the external state. As a result,
it turns out that $W_{\psi \psi}({\bf r})$ solves the two-particle $s$-wave Schr\"odinger equation at zero energy
\begin{align}
\left[- \nabla^2 + \frac{1}{a_0 r}\right] W_{\psi \psi}({\bf r}) = 0 \label{eq:schroedinger}
\end{align}
with boundary condition $W_{\psi \psi}(0) = 1$. The solution of Eq.~\eqref{eq:schroedinger} up to linear order in $r$ is
\begin{align}
W_{\psi \psi}({\bf r}) &= \begin{cases}
1 + \dfrac{r}{2 a_0} + \ldots & \mathrm{(3D)} \\[2ex]
1 + \dfrac{r}{a_0} + \ldots & \mathrm{(2D)}
\end{cases} . \label{eq:cuspcond}
\end{align}
The OPE therefore provides a concise derivation of the intuitive short-distance factorization
\begin{align}
&\lim_{r \to 0} \Psi(- \frac{\mathbf{r}}{2}, \uparrow; \frac{\mathbf{r}}{2}, \downarrow; \ldots) = W_{\psi \psi}({\bf r}) \, \Psi(\mathbf{0}, \uparrow; \mathbf{0}, \downarrow; \ldots)
\label{eq:shortmanyp}
\end{align}
of the many-particle wavefunction~\eqref{eq:wavefunctionfac} if two particles of opposite spin are close to each other. This factorization was previously considered by Lepage for two-particle systems in the context of effective field theories~\cite{lepage97}. Recently, this was used to derive high-momentum tails for the unitary Fermi gas and the Coulomb gas~\cite{bogner12}.

Focusing on pure states, e.g., the ground state of jellium, the result~\eqref{eq:shortmanyp} together with the fact that the two-particle wave function $W_{\psi \psi}({\bf r})$ is non-analytic at short distances implies power-law tails in the ground state momentum distribution and the
static structure factor. Indeed, for a pure state, the dimensionless and intensive momentum distribution is given by
\begin{align}
&n(\mathbf{q}) = \frac{N}{V} \int d^dR \, \int d^dr \, \prod_{l=2}^{N} d^dr_l \, e^{- i \mathbf{q} \cdot \mathbf{r}} \nonumber \\
&\times \Psi^*({\bf R} - \frac{\mathbf{r}}{2}, \mathbf{r}_2, \ldots, \mathbf{r}_{N})\, \Psi({\bf R} +\frac{\mathbf{r}}{2}, \mathbf{r}_2, \ldots, \mathbf{r}_{N}).
\label{eq:firstqun}
\end{align}
Its asymptotic behavior for large momentum is determined by the integration regions in which both $\mathbf{R} + \mathbf{r}/2$ and $\mathbf{R} - \mathbf{r}/2$ approach one of the particle coordinates $\mathbf{r}_l$ for $l=2,\ldots N$ {\it simultaneously}. By substituting the result~\eqref{eq:shortmanyp} in Eq.~\eqref{eq:firstqun}, the resulting high-momentum tail of the momentum distribution turns out to be given by
\begin{align}\label{eq:resmomdist}
n(\mathbf{q}) =
\begin{cases}
\left(\dfrac{4 \pi}{a_0}\right)^2 \dfrac{n^2 g(0)}{q^8} + \ldots & \mathrm{(3D)}  \\[2ex]
\left(\dfrac{2 \pi}{a_0}\right)^2 \dfrac{n^2 g(0)}{q^6} + \ldots & \mathrm{(2D)}
\end{cases} ,
\end{align}
in accordance with the results in Refs.~\cite{kimball73,kimball75, rajagopal77, bogner12}.

The pair distribution function in first-quantized form reads (specifying to the translation invariant case)
\begin{align}
n^2 g(\mathbf{r}) &= N (N-1) \, \int \prod_{l=3}^{N} d^dr_l  | \Psi( \mathbf{0}, \mathbf{r}, \mathbf{r}_3, \ldots, \mathbf{r}_{N}) |^2 . \label{eq:firstqug}
\end{align}
Inserting Eq.~\eqref{eq:shortmanyp} in the definition of the pair correlation function, Eq.~\eqref{eq:firstqug}, we obtain
\begin{align}
g(r)  &=
\begin{cases}
\left( 1 + \dfrac{r}{a_0} \right)  \, g(0) + \ldots & \mathrm{(3D)}  \\[2ex]
\left( 1 + \dfrac{2 r}{a_0} \right) \, g(0) + \ldots & \mathrm{(2D)}
\end{cases} . \label{eq:OPE-dd-correlator}
\end{align}
The pair distribution function at short distance therefore exhibits a dip, rising linearly with slope $g(0)/a_0$. For an {\it attractive} Coulomb force, where $a_0\to -a_0$, the dip is replaced by a cusp.   The result~\eqref{eq:OPE-dd-correlator} agrees with the one obtained previously by
Kimball~\cite{kimball73,kimball75, rajagopal77}. Following Rajagopal \textit{et al.}~\cite{rajagopal78}, we call this the cusp condition. It is interesting to note that a result which is completely analogous to that in Eq.~\eqref{eq:OPE-dd-correlator} holds for Fermi systems with zero range interactions in one dimension, with the 1D scattering length $a_1$ replacing the Bohr radius~\cite{barth11}. Moreover, it is important to mention that various approximate schemes which have been developed to treat the jellium problem at least in its Fermi liquid phase in fact violate the cusp condition~\eqref{eq:OPE-dd-correlator}. This is true in particular for the standard RPA approximation, which is exact in the long wavelength limit but violates the cusp condition, see e.g. Ref.~\cite{niklasson74}. It is obeyed within extensions of the RPA which include local field corrections like the one developed by Singwi, Tosi, Land and Sj\"olander~\cite{singwi68}, at the expense, however, of violating the compressibility sum rule $S(\mathbf{q})\sim q^2$ at long wavelengths $\mathbf{q}\to 0$~\cite{giuliani05,singwi81}.

The short-distance nonanalyticity in the pair distribution function leads to an asymptotic power law in the static structure factor
\begin{align}
S({\bf q}) &= 1 + n \int d^dr \, e^{- i {\bf q} \cdot {\bf r}} \, \left(g(r) - 1\right) .
\end{align}
Substituting Eq.~\eqref{eq:OPE-dd-correlator}, the static structure factor behaves like
\begin{align}
S(\mathbf{q}) -1  &=
\begin{cases}
-\dfrac{8\pi}{a_0} \dfrac{n g(0)}{q^4} & \mathrm{(3D)} \\[2ex]
- \dfrac{4\pi}{a_0} \dfrac{n g(0)}{q^3} & \mathrm{(2D)}
\end{cases} ,
\label{eq:SF-high-momentum}
\end{align}
at large momentum, where we have used that the Fourier transform of $r$ in three and two dimensions is $- 8\pi / q^4$ and $-2\pi / q^3$, respectively, in the sense of distributions. The results in Eq.~\eqref{eq:SF-high-momentum} are again in accordance with those obtained in Refs.~\cite{kimball73,kimball75, rajagopal77, bogner12}.

The tails~\eqref{eq:resmomdist} and~\eqref{eq:SF-high-momentum} in both the momentum distribution and the static structure factor are present for any state of the Coulomb system, a property that will be derived in detail in the following section. Here, we focus on the particular case of zero temperature and moderate values of $r_s$, where the jellium ground state is a uniform Fermi liquid. The characteristic momentum scale beyond which the asymptotic behavior applies is then set by the Fermi wave vector $k_F$, which is related to the average interparticle distance $r_0$ via $k_F=1/(\alpha r_0)$, with $\alpha=(4/9\pi)^{1/3}\simeq 0.521$ in 3D and $\alpha=1/\sqrt{2}\simeq 0.707$ in 2D. Introducing a dimensionless strength $s$ of the tail in the momentum distribution via $n_{\sigma}(\mathbf{q}) \to s\, (k_F/q)^{2d+2}$ one obtains~\cite{gori02}
\begin{align}
s(r_s)  &=
\begin{cases}
\dfrac{9}{2}\alpha^8\, g(0) r_s^2 & \mathrm{(3D)}  \\[2ex]
2 \alpha^6\, g(0) r_s^2 & \mathrm{(2D)}
\end{cases} . \label{eq:tailstrength}
\end{align}

The dimensionless strength $s$ is a continuous function of $r_s$, vanishing in both limits $r_s\to 0$ and $r_s\to\infty$. Indeed, the power-law tail in the momentum distribution is present even in the Wigner crystal, as long as the spin-polarization $\zeta$ remains less than one (recall Eq.~\eqref{eq:g_2onsite}). The contact in this limit is expected to vanish in an exponential manner with $r_s$. In the opposite limit of high density, the fact that $g(0) = 1/2 + {\cal O}(r_s)$~\cite{kimball76} yields $s\sim r_s^2$. The function $s(r_s)$, therefore, must have a maximum, whose value appears to be much smaller than one. Indeed, according to recent quantum Monte Carlo calculations of the momentum distribution in the Fermi liquid phase of the jellium model in 2D~\cite{drummond08} and 3D~\cite{holzmann11}, the resulting dimensionless strengths $s(10)\simeq 0.006$ (2D) or $\simeq 0.009$ (3D) of the power law tail are surprisingly small even at $r_s=10$. In particular, they are almost two orders of magnitude smaller than the corresponding value $s(\infty)=32 \ln{2}/(3\pi^2)\simeq 0.749$ of a Fermi gas with infinite short range repulsion in 1D~\cite{barth11}.

In contrast to zero range interactions, where the value of the contact determines the complete thermodynamics by a simple coupling constant integration~\cite{tan05}, the ground state energy of the jellium model requires knowledge of the pair distribution function at {\it all} distances. Interpolation schemes for the static structure factor and thus the complete pair distribution function $g(r)$ which properly account for both the long- and short-distance behavior of the homogeneous, unpolarized electron gas have been proposed by Gori-Giorgi, Sacchetti and Bachelet~\cite{gori00} and may be used to develop improved versions of the  exchange and correlation energy functionals in density functional theory~\cite{gori06,attaccalite02}.

\section{DIRECT OPE OF THE CORRELATORS}\label{sec:asymptotics}

In the following, we will show how the OPE can be used to perform an expansion of the one-particle density matrix and the pair distribution function~\eqref{eq:onepdm-def} and~\eqref{eq:PD-definition} at the operator level. Apart from providing an alternative derivation of the high-momentum tails which avoids discussing the many-particle wavefunction, this method makes evident a point stressed already in our introduction: the short-distance properties derived here are valid completely independent of the state of the system. In particular, they hold in arbitrary few- or many-body states or in equilibrium at any temperature.

We start by considering the static structure factor, which - for $\mathbf{q}\ne 0$ - is just the Fourier transform
\begin{equation}
S(\mathbf{q})=\frac{1}{N} \int d^dR \int d^dr \, e^{- i \mathbf{q} \cdot \mathbf{r}} \langle n \left(\mathbf{R} - \frac{\mathbf{r}}{2}\right)
n \left(\mathbf{R} + \frac{\mathbf{r}}{2}\right) \rangle
\label{eq:SF-definition}
\end{equation}
of the density correlator. Equations \eqref{eq:PD-definition} and \eqref{eq:SF-definition} imply that its asymptotic behavior for large momentum $q$ is dominated by the short distance behavior of the pair distribution function. The pair distribution function is connected to the $\uparrow\downarrow$ density correlator $\psi_{\uparrow}^\dagger \psi_{\uparrow}^{}  \left(- \frac{\mathbf{r}}{2}\right)  \psi_{\downarrow}^{\dagger} \psi_{\downarrow}^{} \left(\frac{\mathbf{r}}{2}\right)$ via definition \eqref{eq:PD-definition}. As shown in detail in App.~\ref{sec:diagcalc}, the short-distance OPE of this correlator to linear order in $r$ is
\begin{align}
&\psi_{\uparrow}^\dagger \psi_{\uparrow}^{}  \left(- \frac{\mathbf{r}}{2}\right)  \psi_{\downarrow}^{\dagger} \psi_{\downarrow}^{} \left(\frac{\mathbf{r}}{2}\right) \nonumber \\  &= \begin{cases}
\left( 1 + \dfrac{r}{a_0} \right)  \, \psi_\downarrow^\dagger \psi_\uparrow^\dagger \psi_\uparrow^{} \psi_\downarrow^{}(0) + \ldots &\mathrm{(3D)}  \\[2ex]
\left( 1 + \dfrac{2 r}{a_0} \right) \, \psi_\downarrow^\dagger \psi_\uparrow^\dagger \psi_\uparrow^{} \psi_\downarrow^{}(0) + \ldots & \mathrm{(2D)}
\end{cases} \label{eq:dd-correlator-OPE} ,
\end{align}
where we have omitted the analytic term of order ${\bf r}$ since it does not contribute to the high-momentum asymptotics. At this order, additional four-fermion operators involving only one particle species do not contribute because of the anticommutation relations obeyed by the fermion fields. The OPE \eqref{eq:dd-correlator-OPE}, together with the definitions of the pair correlation function \eqref{eq:PD-definition} and the static structure factor \eqref{eq:SF-definition}, reproduces the high-momentum behavior \eqref{eq:SF-high-momentum}. In particular, when taking the expectation value of \eqref{eq:dd-correlator-OPE}, the contact operator $\psi_\downarrow^\dagger \psi_\uparrow^\dagger \psi_\uparrow^{} \psi_\downarrow^{}(0)$ produces the pair correlation function at zero separation.

The momentum distribution $n_{\sigma} (\mathbf{q})$ describes the probability to find a particle of spin $\sigma$ with momentum $\mathbf{q}$. In second quantization, it is defined as the Fourier transform of the one-particle density matrix:
\begin{align}
&n_\sigma(\mathbf{q}) = \dfrac{1}{V} \int d^dx \int d^dy \, e^{- i {\bf q} \cdot ({\bf y} - {\bf x})} \, \langle \gamma^{(1)}_\sigma({\bf x}, {\bf y}) \rangle \nonumber \\
&= \dfrac{1}{V} \int d^dR \int d^dr \, e^{- i \mathbf{q} \cdot \mathbf{r}} \langle \psi_\sigma^\dagger\left(\mathbf{R} - \frac{\mathbf{r}}{2}\right) \psi_\sigma^{}\left(\mathbf{R} + \frac{\mathbf{r}}{2}\right) \rangle . \label{eq:momentumdist}
\end{align}
The non-analytic Wilson coefficients in a short-distance expansion of the one-particle density matrix therefore determine the high momentum tail of $n_\sigma({\bf q})$.  A quite elaborate calculation, which is discussed in detail in App.~\ref{sec:diagcalc}, shows that the OPE of the one-particle density matrix is given by
\begin{widetext}
\begin{align}
\psi_\sigma^\dagger\left(- \frac{\mathbf{r}}{2}\right) \psi_\sigma^{}\left(\frac{\mathbf{r}}{2}\right)  &=
\begin{cases}
\left[e^{- \tfrac{\mathbf{r}}{2} \cdot \nabla} \psi_\sigma^\dagger\left(0\right)\right]   \left[e^{\tfrac{\mathbf{r}}{2} \cdot \nabla} \psi_\sigma^{}\left(0\right)\right] - \dfrac{r^5}{2880 \pi} \left(\dfrac{4 \pi}{a_0}\right)^2 \, \psi_\downarrow^\dagger \psi_\uparrow^\dagger \psi_\uparrow^{} \psi_\downarrow^{}(0) + \ldots & \mathrm{(3D)}  \\[2ex]
\left[e^{- \tfrac{\mathbf{r}}{2} \cdot \nabla} \psi_\sigma^\dagger\left(0\right)\right] \left[e^{\tfrac{{\bf r}}{2} \cdot \nabla} \psi_\sigma^{}\left(0\right)\right] - \dfrac{r^4 \log r}{128 \pi} \left(\dfrac{2 \pi}{a_0}\right)^2 \, \psi_\downarrow^\dagger \psi_\uparrow^\dagger \psi_\uparrow^{} \psi_\downarrow^{}(0) + \ldots & \mathrm{(2D)}
\end{cases} .
\label{eq:opemomentum}
\end{align}
\end{widetext}
The first Wilson coefficients of the bilinear operators are the coefficients in a Taylor expansion of the operators on the left-hand side. The contact operator ${\cal O}_c$ defined in Eq.~\eqref{eq:contact} is the leading order term associated with a non-analytic Wilson coefficient of order $\mathcal{O}(r^5)$ and $\mathcal{O}(r^4 \log r)$ in 3D and 2D, respectively. Substituting Eq.~\eqref{eq:opemomentum} in~\eqref{eq:momentumdist}, we obtain
\begin{align}\label{eq:resmomdist-imbalanced}
n_{\sigma}(\mathbf{q}) =
\begin{cases}
\left(\dfrac{4 \pi}{a_0}\right)^2 \dfrac{ n_{\uparrow} n_{\downarrow} g_{\uparrow \downarrow}(0)}{q^8} + \ldots & \mathrm{(3D)}  \\[2ex]
\left(\dfrac{2 \pi}{a_0}\right)^2 \dfrac{n_{\uparrow} n_{\downarrow} g_{\uparrow \downarrow}(0)}{q^6} + \ldots & \mathrm{(2D)}
\end{cases}
\end{align}
for an arbitrary state with a possible non-vanishing spin polarization $\zeta$. Summing over $\sigma = \uparrow,\downarrow$, we recover the previous result for the momentum distribution at large momentum in the spin-balanced Coulomb gas, Eq.~\eqref{eq:resmomdist}. The fact that the non-analytic terms in Eq.~\eqref{eq:opemomentum} appear at the level of operators shows that the tails in the momentum distribution are also present in phases where translation invariance is broken, in a few-body situation, or at arbitrary temperatures. In a Wigner crystal, for instance, the product $n_{\uparrow} n_{\downarrow} g_{\uparrow \downarrow}(0)$ has to be replaced by
\begin{equation}
\frac{1}{V} \int d^dR \, \langle \psi_\downarrow^\dagger \psi_\uparrow^\dagger \psi_\uparrow^{} \psi_\downarrow^{}(\mathbf{R}) \rangle\, ,
\end{equation}
which is again an intensive quantity in the thermodynamic limit $N,V\to\infty$ at fixed average densities $n_{\uparrow}, n_{\downarrow}$. In a few-body situation, in turn, these densities vanish but there is still a finite expectation value of the contact operator. For the hydrogen atom for instance, one finds
\begin{equation}
\int d^3 R \, \langle \psi_\downarrow^\dagger \psi_\uparrow^\dagger \psi_\uparrow^{} \psi_\downarrow^{}(\mathbf{R}) \rangle= |\psi_{n00}(0)|^2 ,
\end{equation}
in agreement with the result derived in the introduction.

\section{HIGH-TEMPERATURE VERSUS CLASSICAL LIMIT}\label{sec:hight}

Beyond the derivation of exact relations which constrain the short-distance properties of Coulomb systems in quite general terms and which - as has been shown in the preceding sections - all involve the contact $\langle {\cal O}_c(\mathbf{R}) \rangle$, quantitative results for specific phases of jellium or non-trivial few-body Coulomb systems require to calculate the value of the contact as a function of both the interaction strength $r_s$ and temperature $T$.  Since relevant values of $r_s$ are beyond the regime where perturbation theory can be applied, this can only be achieved numerically, for instance via quantum Monte Carlo calculations, see e.g. Refs.~\cite{drummond08,holzmann11} for some recent results. In the following, we calculate the value of the contact in the classical and the high-temperature limit. Surprisingly, it turns out that for Coulomb interactions these two limits are not equivalent. In fact, they turn out to be completely opposite.

Consider the Coulomb gas in the regime
\begin{align}
k_BT \gg \frac{\hbar^2 n^{2/d}}{m} ,
\end{align}
where the thermal energy is much larger than the degeneracy energy. This is the standard limit of a non-degenerate gas, in which the average interparticle spacing $n^{-1/d}$ is much larger than the thermal wavelength $\lambda_T = \hbar (2 \pi/mk_BT)^{1/2}$:
\begin{align}
n^{1/d} \lambda_T \ll 1 .
\label{eq:virial}
\end{align}
In this limit, thermodynamic properties can be calculated by expanding in powers of the fugacity $z=\exp{(\beta\mu)}=n\lambda_T^d/2\ll 1$~\cite{footnote4}. The non-degeneracy condition~\eqref{eq:virial} does not involve the strength $e^2$ of the interaction and is satisfied both in the infinite temperature and in the classical limit. Now, for systems with Coulomb interactions, there is a second and purely classical, so-called Bjerrum length $\ell_B = e^2/(k_BT)$, which - keeping $\hbar$ finite - eventually becomes shorter than the thermal length at sufficiently high temperatures. As a result -- already noted in Ref.~\cite{schweng91} -- the order in which the limits $T\to\infty$ or $\hbar\to 0$ is taken matters. Taking $T \to \infty$ {\it before} $\hbar \to 0$ results in the following hierarchy of length scales:
\begin{align}
n^{-1/d} \gg a_0 \gg \lambda_T \gg \ell_B .
\end{align}
In turn, taking the classical limit $\hbar \to 0$ before $T \to \infty$, we find:
\begin{align}
n^{-1/d} \gg \ell_B \gg \lambda_T \gg a_0 .
\end{align}
As will be shown below, these two limits give quite different results for the value of the contact. Since the gas is non-degenerate in both cases, the contact value of the pair distribution function can be calculated to leading order in the virial expansion, which just involves an integration
\begin{align}
n^2 g(0) &= z^2 \frac{2^{d/2}}{\lambda_T^d} \int \frac{d^dp}{(2\pi\hbar)^d} \, e^{- \beta p^2/m} |\psi_p(0)|^2 \, ,
\label{eq:virial_g(0)}
\end{align}
of the square of the relative Coulomb wavefunction  $|\psi_p(0)|^2$ at the origin with the classical Boltzmann distribution for the relative momentum $p$. Here, $z = e^{\beta \mu}$ is the fugacity while $|\psi_p(0)|^2$ is given by~\cite{gottfried66}
\begin{align}
|\psi_p(0)|^2 &= \begin{cases}
 	\dfrac{2 \pi \xi}{e^{2 \pi \xi} - 1} & \mathrm{(3D)} \\[2ex]
 	\dfrac{2}{e^{2 \pi \xi} + 1} & \mathrm{(2D)}
\end{cases} ,
\end{align}
with $\xi = m e^2/2 \hbar p$. In the infinite temperature limit, the integration gives
\begin{align}
g(0) &= \begin{cases}
\displaystyle\frac{1}{2} \displaystyle\left(1 - \sqrt{2} \pi \dfrac{\ell_B}{\lambda_T} + \ldots\displaystyle\right) & \mathrm{(3D)} \\[2ex]
\displaystyle\frac{1}{2} \displaystyle\left(1 - \dfrac{\pi^2}{\sqrt{2}} \dfrac{\ell_B}{\lambda_T} + \ldots\displaystyle\right) & \mathrm{(2D)}
\end{cases} . \label{eq:virialtemp}
\end{align}
The leading order term is the expected result for a classical ideal gas. The corrections involve the ratio $\ell_B/\lambda_T$ and thus vanish for large temperatures like $\sim 1/\sqrt{T}$. Note that, although $g(0)$ is of order one in this limit, the high-momentum tails in this limit are present only for very large $q \gg \lambda_T^{-1}$.

More relevant for low temperature, non-degenerate plasmas is the classical limit $\hbar \to 0$ for which Ry$\gg k_BT$. This limit is reached, for example, for Coulomb gases of charged dust particles in astrophysics~\cite{ivlev12}. In this limit the integration in~\eqref{eq:virial_g(0)} gives rise to a contact which vanishes exponentially like
\begin{align}
g(0) &= \begin{cases}
  \dfrac{4 \pi^2 2^{1/3}}{3^{1/2}} \left(\dfrac{\ell_B}{\lambda_T}\right)^{4/3} \, e^{- \tfrac{3 \pi}{2^{1/3}} \left(\tfrac{\ell_B}{\lambda_T}\right)^{2/3}} + \ldots & \mathrm{(3D)} \\
\dfrac{2 \pi 2^{1/3}}{3^{1/2}} \left(\dfrac{\ell_B}{\lambda_T}\right)^{1/3} \, e^{- \tfrac{3 \pi}{2^{1/3}} \left(\tfrac{\ell_B}{\lambda_T}\right)^{2/3}} + \ldots & \mathrm{(2D)}
\end{cases} . \label{eq:virialclass}
\end{align}
Equations~\eqref{eq:virialtemp} and~\eqref{eq:virialclass} are the main results of this section. The Wilson coefficients of the momentum distribution and the static structure factor diverge as $1/a_0^2 \sim 1/\hbar^4$ and $1/a_0 \sim 1/\hbar^2$  as $\hbar \to 0$. The high momentum tails are expected to occur for $q \gg \lambda_T^{-1}$. Thus, their characteristic momentum scale is pushed to infinity as $\hbar \rightarrow 0$. The total weight of the tails, containing both the Wilson coefficients and $g(0)$, is exponentially suppressed and guarantees a well defined kinetic energy in this limit. We point out that different high-momentum tails are present in the classical system. Indeed, the structure factor of the classical electron gas decreases as $1/q^2$ at intermediate momentum $\kappa < q < 1/l_B$, where $\kappa = \sqrt{n l_B}$ is the inverse Debye-H\"uckel length~\cite{berggren69}.

\subsection{Diagrammatic derivation}

In the infinite temperature limit, the virial expansion of the contact $g(0)$ in Eq.~\eqref{eq:virialtemp} can also be obtained using a diagrammatic formalism. This method was originally introduced by Vedenov and Larkin to derive the equation of state of an electron gas~\cite{Vedenov59}, and was recently used by various groups to determine the virial expansion of a Fermi gas with short-range interactions~\cite{bedaque03,rupak07,kaplan11,leyronas11}.

The perturbation series of an arbitrary correlator involves all Feynman diagrams that connect to the operator insertions. In the absence of a small parameter, this gives rise to a very large number of diagrams. The key point of the method is that in the infinite temperature limit, this number is drastically reduced by exploiting the causal structure of the propagators, which are defined as
\begin{align}
G(\tau, {\bf k}) &= \begin{cases}
    	- (1 - n_{\bf k}) e^{- (\varepsilon_{\bf k} - \mu) \tau} & \tau > 0 \\
    	n_{\bf k} e^{- (\varepsilon_{\bf k} - \mu) \tau} & \tau < 0
   	\end{cases},
\end{align}
where $n_{\bf k} = 1/(e^{\beta (\varepsilon_{\bf k} - \mu)} + 1)$ is the Fermi-Dirac distribution. In imaginary time, they carry the whole dependence on the fugacity and, thus, an expansion of a diagram in the fugacity corresponds to an expansion of the propagators:
\begin{align}
G(\tau, {\bf k}) = G^{(0)}(\tau, {\bf k}) + G^{(1)}(\tau, {\bf k}) + {\cal O}(z^2) ,
\end{align}
with
\begin{align}
G^{(0)}(\tau, {\bf k}) &= - \Theta(\tau) \, e^{\mu \tau} e^{- \varepsilon_{\bf k} \tau} \quad {\rm and} \label{eq:barepropLO} \\
G^{(1)}(\tau, {\bf k}) &= z \, e^{\mu \tau} e^{- \varepsilon_{\bf k} (\beta + \tau)} .
\end{align}
$\Theta$ is the Heaviside step function. The coefficients $G^{(i)}$ with $i=1,2,\ldots$ can be treated as separate diagrammatic elements, e.g. by denoting them by a line that is slashed $i$ times. The order of a diagram in $z$ is then set by the sum $\sum_i i N_i$, where $N_i$ is the number of propagators of type $i$. It is important to note that the leading order term~\eqref{eq:barepropLO} is purely retarded, and while the calculation of a general order may be unwieldy, only a very limited number of diagrams contribute to the virial expansion of a correlator to leading order in $z$.

\begin{figure}[t!]
\begin{center}
\subfigure[]{\scalebox{0.5}{\epsfig{file=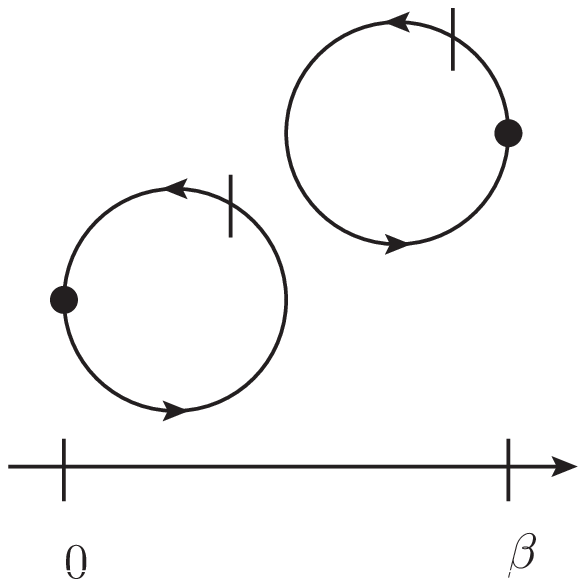,bb=200 551 395 721}}\label{fig:2a}} \hspace*{0.5cm}
\subfigure[]{\scalebox{0.5}{\epsfig{file=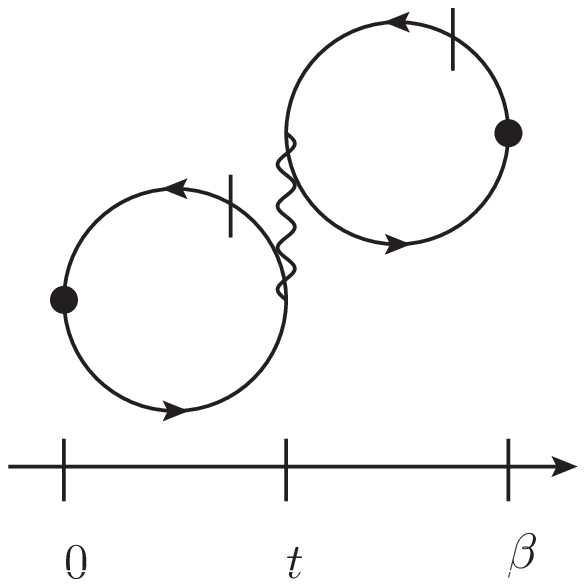,bb=200 551 395 721}}\label{fig:2b}}
\caption{(a) ${\cal O}(e^0)$ and (b) ${\cal O}(e^2)$ contribution to $g(0)$. Imaginary time runs from the left to the right. Black dots denote the density operator, the wavy line the instantaneous Coulomb interaction, and (slashed) lines the particle propagator as explained in the text.}
\label{fig:2}
\end{center}
\end{figure}

Figure~\ref{fig:2} shows the two leading order contribution in $e^2$ to the pair correlation function
\begin{align}
g_{\uparrow\downarrow}(0) &= \frac{1}{n_\uparrow n_\downarrow} \left\langle T_\tau n_{\uparrow}(\beta) n_{\downarrow} (0) \right\rangle ,
\end{align}
where we define the density operator in the usual sense $n_\sigma = - \lim_{\tau \to 0-} T_\tau \psi_\sigma(\tau) \psi_\sigma^\dagger(0)$ to avoid an ordering ambiguity~\cite{fetter71}. Figure~\ref{fig:2a} corresponds to the noninteracting result
\begin{align}
g_{\uparrow\downarrow}^a(0) &= 1 ,
\end{align}
where we use the relation $n_\sigma = z_\sigma \lambda_T^{-d}$. The ${\cal O}(e^2)$ contribution in Fig.~\ref{fig:2b} reads
\begin{align}
g_{\uparrow\downarrow}^b(0) &= \begin{cases}
-\sqrt{2} \pi \dfrac{\ell_B}{\lambda_T} & \mathrm{(3D)} \\[2ex]
- \dfrac{\pi^2}{\sqrt{2}} \dfrac{\ell_B}{\lambda_T} & \mathrm{(2D)}
\end{cases} ,
\end{align}
which coincides with our previous result~\eqref{eq:virialtemp}. Note that higher order contributions to $g(0)$ contain infrared divergences. They can be removed by summing the divergent parts of an infinite number of ring diagrams~\cite{Vedenov59}, which gives rise to Debye-H{\"u}ckel corrections that are of higher order in the density.

\section{SUMMARY AND CONCLUSIONS}\label{sec:summary}

In summary, we used a short-distance operator product expansion to derive the high-momentum tail of the structure factor and the momentum distribution of the Coulomb gas. Since these results are based on operator identities, they hold for pure states as well as for mixtures, and, in particular, for different phases, such as a Fermi liquid or a Wigner crystal. The key idea behind the derivation of our exact results is the separation of short and long distance scales. The functional dependence on the high-energy scales can be calculated exactly while the low-energy contribution factorizes. This multiplicative constant is the contact value of the pair distribution function, consistent with the intuitive expectation that the short-distance physics is determined by the probability to find two particles at the same point. These results are in close analogy to the Tan relations for zero range interactions~\cite{tan05}, in which an analogous contact enters the coefficients of the high-momentum tails.

Furthermore, we calculated the contact in explicit form for non-degenerate Coulomb gases, using a virial expansion. It turns out that there are two possible limits which yield quite different results: in the high-temperature limit, the contact approaches the ideal Fermi gas value with power-law corrections in the temperature. By contrast, in the classical limit, the contact vanishes exponentially as $\hbar\to 0$, a behavior which is crucial to ensure a well-defined transition to the classical regime in which the Coulomb repulsion between the particles prevents them from being at coincident points.

The universal relations obtained in the present article are by far not exhaustive. Indeed, many more relations could be derived within the framework introduced here. A short-time OPE analogous to Refs.~\cite{son10,hofmann11,nishida12} would give results for dynamical correlators, such as, for example, the current response function or the dynamic structure factor, which display short-range correlations that are not captured in a random phase approximation~\cite{sternemann05}. A similar analysis can be carried out for the spectral function, which possesses a high-frequency tail as derived in~\cite{glick71,pavlyukh12}. Beyond applying the OPE to the Coulomb gas, it should be straightforward to generalize the results in this article to other many-fermion systems with long-range interactions, such as quantum gases of dipolar particles, which have recently been studied experimentally~\cite{griesmaier05,aikawa12,lu12}.

\section*{Acknowledgments}

We thank Tilman Enss, Markus Holzmann, Hartmut L\"owen, Dietrich Roscher, Richard Schmidt and Matthew Wingate for discussions. JH is supported by CHESS, STFC, St. John's College, Cambridge, and by the Studienstiftung des deutschen Volkes. MB is supported
by the DFG research unit ``Strong Correlations in Multiflavor Ultracold Quantum Gases''.

\appendix
\raggedbottom

\section{Feynman Rules}\label{sec:feyrules}

This appendix summarizes the Feynman rules of the Hamiltonian~\eqref{eq:Hamiltonian} in momentum space. An energy $\omega$ and a momentum $\mathbf{q}$ are assigned to each internal and external line. The bare propagator is denoted by a continuous line and contributes a factor $G_0(\omega, {\bf q}) = i/(\omega-q^2/2m+i0)$. We represent the interaction between two fermions by a wavy line. It contributes a factor $4 \pi i e^2/q^2$ in 3D and $2 \pi i e^2/q$ in 2D, where q is the difference between the center of mass momenta of the ingoing and outgoing fermions. Finally, each undetermined momentum and energy is integrated with measure $\int d^dq/(2\pi)^d \int d\omega/2\pi$.

\begin{figure*}[t!]
\begin{center}
\subfigure[]{\scalebox{0.6}{\epsfig{file=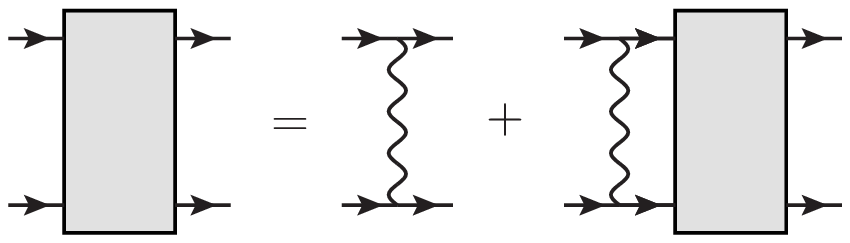,bb=169 646 426 721}}\label{fig:3a}}
\subfigure[]{\scalebox{0.6}{\epsfig{file=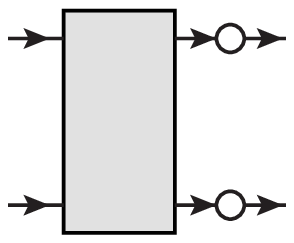,bb=249 646 346 721}}\label{fig:3b}}
\subfigure[]{\scalebox{0.6}{\epsfig{file=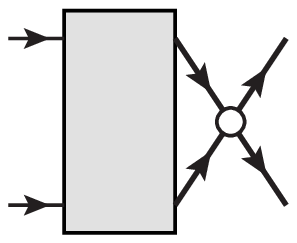,bb=249 646 347 721}}\label{fig:3c}}
\subfigure[]{\scalebox{0.6}{\epsfig{file=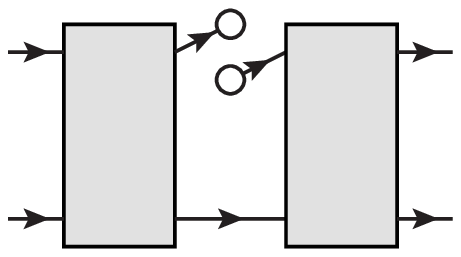,bb=225 642 370 721}}\label{fig:3d}}
\subfigure[]{\scalebox{0.6}{\epsfig{file=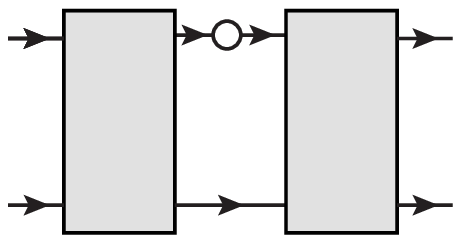,bb=225 646 370 721}}\label{fig:3e}}
\caption{(a) Bethe-Salpeter equation for the T-matrix, which is indicated by a gray rectangle. (b-e) Contribution to the two-particle matrix elements of (b) the density-density correlator $n_\uparrow(- \tfrac{\mathbf{r}}{2}) n_\downarrow(\tfrac{\mathbf{r}}{2})$, (c) the contact of the pair distribution function $g(0)$, (d) the one-particle density matrix $\psi_\sigma^\dagger(- \tfrac{\mathbf{r}}{2}) \psi_\sigma(\tfrac{\mathbf{r}}{2})$, and (e) bilinear operators, respectively.}
\label{fig:3}
\end{center}
\end{figure*}

The T-matrix insertion $i T({\bf p}, {\bf p}', k)$ is denoted by a gray rectangle, where ${\bf p}$ and ${\bf p}'$ are the center of mass momentum of the ingoing and outgoing atoms, respectively. We denote the center of mass energy of the ingoing atoms by $E = k^2/m$. Note that this energy is not necessarily on-shell, i.e. we do not impose the condition $k^2 = p^2 = p'^2$. The T-matrix solves the Bethe-Salpeter equation, depicted diagrammatically in Fig.~\ref{fig:3a}, and reads~\cite{schwinger64,holstein94,dittrich99,chen72}
\begin{align}
&i T(\mathbf{p}, \mathbf{p}', k) \nonumber \\
&=
\begin{cases}
- 16 i \pi e^2 k^2 \displaystyle\int_0^1 dx \, x^{i\xi} \dfrac{d}{dx} \frac{x}{H({\bf p},{\bf p}',k)}  & \mathrm{(3D)}  \\[3ex]
- 4 i \pi e^2 k \displaystyle\int_0^1 dx \, x^{i\xi} \dfrac{d}{dx} \frac{x^{1/2}}{H^{1/2}({\bf p},{\bf p}',k)} & \mathrm{(2D)}
\end{cases} ,
\end{align}
where $H({\bf p},{\bf p}',k) = 4 k^2 (\mathbf{p} - \mathbf{p}')^2 x - (k^2 - p^2) (k^2 - p'^2) (1-x)^2$ and $\xi = m e^2/2 k$ is called the Sommerfeld parameter.

\section{Operator Product Expansion} \label{sec:diagcalc}

In this appendix, we collect the matrix elements and momentum integrals needed to perform the OPE for the Coulomb gas. For simplicity, we set $\hbar = 1$ in the following. Since the operator product expansion is state-independent, it is sufficient to evaluate the matrix elements of the operator products~\eqref{eq:onepdm-def} and~\eqref{eq:PD-definition} and the local operators between selected few-particle states. The Wilson coefficients are determined by matching the terms in an expansion of these expectation values in the external parameters of the state. The Wilson coefficients of the bilinear operators, i.e., operators that contain one field operator and its hermitian conjugate, are the coefficients in a Taylor expansion of the operator product, which can be obtained by matching the matrix elements between a one-particle state. To compute the contact's Wilson coefficient, we choose a two-particle state with zero relative momentum and (off-shell) energy $k^2/m$, which we denote by $\left< k | \mathcal{O} | k \right>$. These matrix elements are the sum of four diagrams with either scattering or no scattering in the initial and final states. Some of them are depicted in Fig.~\ref{fig:3}.

\subsubsection{Structure factor}

As explained in Sec.~\ref{sec:asymptotics}, the short-distance behavior of the pair correlation function is inferred from an OPE of the $\uparrow\downarrow$ density correlator $n_\uparrow(- \tfrac{\mathbf{r}}{2}) n_\downarrow(\tfrac{\mathbf{r}}{2})$. The matrix element of this operator between a two-particle state can be expressed in terms of diagram \ref{fig:3b}. The complete matrix element is the sum of four diagrams with either scattering or no scattering in the initial and final states:
\begin{widetext}
\begin{align} \label{eq:SF-fullme}
\langle k | n_\uparrow(- \frac{\mathbf{r}}{2}) n_\downarrow(\frac{\mathbf{r}}{2}) | k \rangle &= \left[1 + \int \frac{d\omega}{2\pi} \int \frac{d^dq}{(2 \pi)^d} e^{i \mathbf{q} \cdot \mathbf{r}} i T (\mathbf{0},\mathbf{q},k) G_0 (\omega, \mathbf{q}) G_0 (E -\omega, -\mathbf{q})\right]^2 \nonumber \\
&=
\begin{cases}
\displaystyle \left(1 + \frac{r}{a_0} \right) \left[ 4 \int_0^1 dx \, x^{i\xi} \frac{d}{dx} \frac{x}{(1+x)^2} \right]^2 + \mathcal{O} (r^2)  & \mathrm{(3D)}  \\[2ex]
\displaystyle \left(1 + \frac{2 r}{a_0} \right) \left[ 2 \int_0^1 dx \, x^{i\xi} \frac{d}{dx} \frac{x^{1/2}}{1+x} \right]^2 + \mathcal{O} (r^2)   & \mathrm{(2D),}
\end{cases}
\end{align}
where we used the integrals
\begin{align}
& \int \frac{d^3q}{(2\pi)^3} \, \frac{e^{i {\bf q} \cdot {\bf r}}}{(q^2 - a^2) (q^2 - b^2)}
=  \frac{i}{4 \pi} \frac{1}{a+b} - \frac{r}{8 \pi} + {\cal O}(r^2)
\label{eq:SF-integrals1}
\end{align}
and
\begin{align}
& \int \frac{d^2q}{(2\pi)^2} \, \frac{e^{i {\bf q} \cdot {\bf r}}}{(q^2 - a^2)^{1/2} (q^2 - b^2)} = \frac{i}{2\pi} \frac{\arccos \tfrac{a}{b}}{\sqrt{b^2-a^2}} - \frac{r}{2\pi} + {\cal O}(r^2) .
\label{eq:SF-integrals2}
\end{align}
The factor in square brackets in Eq.~\eqref{eq:SF-fullme} depends on the details of the states and must not contribute to the Wilson coefficients. It is matched by the expectation value of the contact  $\psi_\downarrow^\dagger \psi_\uparrow^\dagger \psi_\uparrow^{} \psi_\downarrow^{}(0)$. Consider the diagram in Fig. \ref{fig:3c}:
\begin{align}
&\int \frac{d^dq}{(2 \pi)^d} \int \frac{d\omega}{2\pi} \, (i T({\bf 0}, {\bf q}, k)) G_0(\omega, {\bf q}) G_0(E - \omega, - {\bf q}) =
\begin{cases}
-4 i \xi \displaystyle\int_0^1 dx \, \frac{x^{i\xi}}{(1+x)^2} & \mathrm{(3D)} \\[2ex]
-2 i \xi \displaystyle\int_0^1 dx \, \dfrac{x^{i\xi-1/2}}{1+x} & \mathrm{(2D)}
\end{cases} .
\end{align}
The full matrix element of $\psi_\downarrow^\dagger \psi_\uparrow^\dagger \psi_\uparrow^{} \psi_\downarrow^{}(0)$ contains three additional diagrams:
\begin{align}\label{eq:Contact}
&\langle k | \psi_\downarrow^\dagger \psi_\uparrow^\dagger \psi_\uparrow^{} \psi_\downarrow^{}(0) | k \rangle =
\begin{cases}
\left[4 \displaystyle\int_0^1 dx \, x^{i\xi} \, \dfrac{d}{dx} \frac{x}{(1+x)^2}\right]^2 \displaystyle\approx \dfrac{1}{4\xi^4}  & \mathrm{(3D)}  \\[2ex]
\left[2 \displaystyle\int_0^1 dx \, x^{i\xi} \, \dfrac{d}{dx} \frac{x^{1/2}}{1+x}\right]^2 \displaystyle\approx \dfrac{1}{16\xi^4} &  \mathrm{(2D)}
\end{cases} ,
\end{align}
where we have expanded the result to leading order in $k$.

\subsubsection{Momentum distribution} \label{sec:1p-dmatrix}

To obtain the asymptotic form of the momentum distribution~\eqref{eq:momentumdist}, one performs an OPE of the nonlocal operator $\psi_\sigma^\dagger(- \tfrac{\mathbf{r}}{2}) \psi_\sigma(\tfrac{\mathbf{r}}{2})$, whose expectation value gives the one-particle density matrix $\gamma(\tfrac{\mathbf{r}}{2}, -  \tfrac{\mathbf{r}}{2})$, cf. Eq.~\eqref{eq:onepdm-def}. Since insertions of this operator on external legs are matched by bilinear operators, the only relevant diagram that contributes to the Wilson coefficient of the contact operator involves scattering in both initial and final state as shown in Fig.~\ref{fig:3d}. As we are only interested in the leading order non-analyticity of the Wilson coefficient of the zero distance pair correlator, we expand the T-matrix, as well as our diagram, as a power series in $k$ around $k=0$. This procedure introduces infrared divergences, which we regulate by introducing an infrared cutoff $\mu$. The expansion of the T-matrix with respect to $k \sim 1/\xi$ is given by
\begin{align}
i T (\mathbf{0},\mathbf{q},k) =
\begin{cases}
\displaystyle i \frac{2 \pi e^2}{\xi^2 q^2}  = \lim_{\mu \rightarrow 0} i \frac{2\pi e^2}{\xi^2 (q^2+\mu^2)}     & \mathrm{(3D)}  \\[2ex]
\displaystyle i \frac{\pi e^2}{2 \xi^2 q}    = \lim_{\mu \rightarrow 0} i \frac{\pi e^2}{\xi^2 \sqrt{q^2+\mu^2}} & \mathrm{(2D).}
\end{cases}
\end{align}
In addition, we expand our propagators as $1/(q^2-k^2) = 1/(q^2+\mu^2) + \mathcal{O}(k^2)$. In the limit $k \rightarrow 0$, diagram \ref{fig:3d} is given by
\begin{align}
\label{eq:1p-dmatrix-2pstates}
\int \frac{d^dq}{(2 \pi)^d} & \int \frac{d\omega}{2\pi} \,e^{i \mathbf{q} \cdot \mathbf{r}}  (i T({\bf 0}, {\bf q}, k \rightarrow 0))^2 G_0(\omega, {\bf q})^2 G_0(E - \omega, - {\bf q})  \nonumber \\
& =
\begin{cases}
\displaystyle \frac{\pi}{48 a_0^2 \xi^4} \left[ \frac{3}{\mu^5} - \frac{r^2}{2\mu^3} +\frac{r^4}{8\mu}-\frac{r^5}{15} \right] + \mathcal{O}(r^6) & \mathrm{(3D)} \\[2ex]
\displaystyle \frac{\pi}{16 a_0^2 \xi^4}  \left[ \frac{1}{2 \mu ^4} - \frac{r^2}{8 \mu ^2} +\frac{r^4}{32} \left(\frac{3}{4} -  \log \frac{e^{\gamma_E}\mu r}{2} \right)\right] + \mathcal{O} (r^5) & \mathrm{(2D).}
\end{cases}
\end{align}
The matrix elements of the one-particle operators (Fig.~\ref{fig:3e}) match the analytic terms in this expansion. The remainder is of order ${\cal O}(r^5)$ and ${\cal O}(r^4 \log r)$ and is matched by the contact operator:
\begin{align}
&W_{{\cal O}_c}(r) =
\begin{cases}
\displaystyle - \left( \frac{4\pi}{a_0} \right)^2 \frac{r^5}{2880 \pi} + \mathcal{O}(r^6) & \mathrm{(3D)} \\[2ex]
\displaystyle  - \left( \frac{2\pi}{a_0} \right)^2 \frac{r^4}{128 \pi} \left( -\frac{3}{2} + \log \frac{e^{\gamma_E} r}{2} \right) +\mathcal{O}(r^5) & \mathrm{(2D)}
\end{cases} .
\end{align}

\begin{figure*}[t!]
\begin{center}
\subfigure[]{\scalebox{0.7}{\epsfig{file=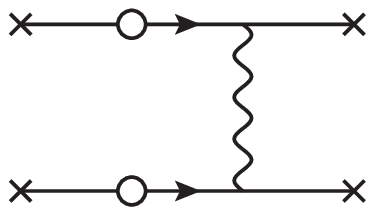,bb=238 654 357 721}}\label{fig:4a}}
\subfigure[]{\scalebox{0.7}{\epsfig{file=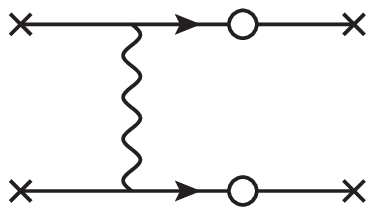,bb=238 654 357 721}}\label{fig:4b}}
\subfigure[]{\scalebox{0.8}{\epsfig{file=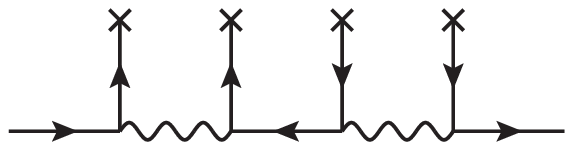,bb=209 673 386 721}}\label{fig:4c}}
\caption{Leading-order contributions to the structure factor (a,b) and momentum distribution (c).}
\label{fig:4}
\end{center}
\end{figure*}
\end{widetext}

\subsubsection{Green's function OPE}

In addition to the derivation outlined in the previous section, the high-momentum tail of the structure factor can also be obtained from a short-time and -distance OPE of the time-ordered density Green's function
\begin{align}
i G_n(\omega, {\bf q}) &= \int dt \int d^dx \, e^{i \omega t - i {\bf q} \cdot {\bf x}} \, \langle T n(t, \mathbf{x}) n(0, \mathbf{0}) \rangle .
\end{align}
For ${\bf q} \neq 0$, it is related to the structure factor by
\begin{align}
S(\mathbf{q}) - 1 &= \frac{1}{n} \lim_{t \to 0-} \int \frac{d\omega}{2 \pi} \, e^{- i \omega t} \, i G_n(\omega, \mathbf{q}) .
\end{align}
The integral is evaluated by closing the contour in a large semicircle in the lower half of the complex $\omega$-plane. Only Wilson coefficients with poles in both half-planes contribute to the high-momentum tail. In the limit of negligible external scales, diagrams have vanishing residue if they can be traversed from one operator insertion to the other by following the fermion lines. The Wilson coefficient of the operator ${\cal O}_c$ are read off directly from the diagrams in Figs.~\ref{fig:4a} and~\ref{fig:4b}:
\begin{align}
&W_{{\cal O}_c}(\omega, {\bf q}) \nonumber \\
&=
\begin{cases}
\dfrac{8 \pi e^2}{q^2} \dfrac{1}{(\omega - \varepsilon_{\bf q} + i0) (- \omega - \varepsilon_{\bf q} + i0)} & \mathrm{(3D)} \\[2ex]
\dfrac{4 \pi e^2}{q^2} \dfrac{1}{(\omega - \varepsilon_{\bf q} + i0) (- \omega - \varepsilon_{\bf q} + i0)} & \mathrm{(2D)}
\end{cases} ,
\end{align}
where $\varepsilon_\mathbf{q} = q^2/2m$, and the external lines couple to the contact $g_{\uparrow\downarrow}(0)$. Performing the contour integration reproduces the result~\eqref{eq:SF-high-momentum}.

We can apply a similar argument to determine the high-momentum tail of the momentum distribution, which is related to the single-particle Green's function $i G_\sigma(t, {\bf x}) = \langle T \psi_\sigma(t, {\bf x}) \psi_\sigma^\dagger(0,{\bf 0}) \rangle$ by~\cite{fetter71}
\begin{align}
n_\sigma(\mathbf{q}) = - \lim_{t \to 0-} \int \frac{d\omega}{2 \pi} \, e^{- i \omega t} \, i G_\sigma({\bf q}, \omega) . \label{eq:nGreen}
\end{align}
This relation was used to derive the momentum distribution of a Fermi gas with short-range interactions~\cite{son10}. The first nonzero contribution is given by the contact operator, which has the Wilson coefficient (cf. Fig.~\ref{fig:4c})
\begin{align}
&W_{{\cal O}_c}(\omega, {\bf q}) \nonumber \\
&=
\begin{cases}
\left(\dfrac{4 \pi e^2}{q^2}\right)^2 \dfrac{- 1}{(\omega - \varepsilon_\mathbf{q} + i0)^2 (- \omega - \varepsilon_\mathbf{q} + i 0)} \quad \mathrm{(3D)} \\[2ex]
\left(\dfrac{2 \pi e^2}{q}\right)^2 \dfrac{- 1}{(\omega - \varepsilon_\mathbf{q} + i0)^2 (- \omega - \varepsilon_\mathbf{q} + i 0)} \quad \mathrm{(2D)}
\end{cases} .
\end{align}
Calculating the residue in Eq.~(\ref{eq:nGreen}) yields the previous result~\eqref{eq:resmomdist-imbalanced}.


\begin{thebibliography}{99}
\bibitem{footnote2}
More precisely, the exact relations from Secs.~\ref{sec:wavefunction} and~\ref{sec:asymptotics} are recovered for the hydrogen atom if we replace $1/a_0 \rightarrow -2/a_0$ in Eqs.~\eqref{eq:resmomdist} and \eqref{eq:SF-high-momentum}. This replacement takes into account the change of the reduced mass from $m$ to $m/2$ as well as the change of the sign of the interaction.
\bibitem{footnote3}
Since the Coulomb potential is long-ranged, the matrix elements reproduce the quantum mechanical result only up to a divergent but $r$-independent phase factor, cf. Refs.~\cite{schwinger64,holstein94}. This does not affect the argument in Sec.~\ref{sec:wavefunction}.
\bibitem{footnote4}
Note that due to the long-range nature of the Coulomb interaction the corrections to this leading order result are non-analytic in the density. For the calculation of the contact at high temperatures or in the classical limit $\hbar\to 0$, however, these corrections are irrelevant.
\end{thebibliography}
\end{document}